\documentclass[12pt,a4paper]{revtex4-2}
\usepackage[utf8x]{inputenc}
\usepackage{ucs}
\usepackage[english]{babel}
\usepackage{amsmath}
\usepackage{amsfonts}
\usepackage{amssymb}
\usepackage{graphicx}
\usepackage[left=2cm,right=2cm,top=2cm,bottom=2cm]{geometry}

\usepackage{bm}

\usepackage{indentfirst}

\usepackage{bbold}

\usepackage{float}
\usepackage[caption = false]{subfig}
\usepackage[OT2,T1]{fontenc}

\DeclareSymbolFont{cyrletters}{OT2}{wncyr}{m}{n}
\DeclareMathSymbol{\Sha}{\mathalpha}{cyrletters}{"58}

\begin{document}

\title{Coherent states in quantum billiards constructed in the basis of the continued eigenfunctions}

\author{I.D. Burkov}
\affiliation{NUST MISIS, Moscow, Russia}

\author{S.S. Seidov}
\affiliation{HSE University, Moscow, Russia, alikseidov@yandex.ru}

\newcommand{\vx}{\bm{x}}
\newcommand{\vp}{\bm{p}}
\newcommand{\vy}{\bm{y}}
\newcommand{\vj}{\bm{j}}
\newcommand{\vk}{\bm{k}}

\newcommand{\bG}{\mathbf{G}}
\newcommand{\bU}{\mathbf{U}}
\newcommand{\bR}{\mathbf{R}}
\newcommand{\bP}{\mathbf{P}}

\renewcommand{\Im}{\operatorname{Im}}
\renewcommand{\Re}{\operatorname{Re}}

\newcommand{\F}{\mathcal{F}}

\begin{abstract}
In  the article a new approach to construction of generalized coherent states in quantum billiards is proposed. The coherent states are defined as the projections of a Gaussian wave function on the basis of the eigenstates of the quantum billiard, continued outside. The continuation is built as the solution of the equivalent Balian--Bloch equation, defined on the entire $\mathbb{R}^2$ and the projection operator is built using the resolvent of the Balian--Bloch equation. In the case of the one--dimensional potential well and billiards, belonging to the Coxeter group, the wave functions of the coherent states were expressed analytically via the Jacobi and Riemann theta functions.
\end{abstract}

\maketitle

\section{Introduction}
Construction of generalized coherent states in quantum billiards is a prominent problem in the field of quantum billiards and quantum chaos \cite{klauder_current_2001, curado_heisenberg-type_2001, curado_generalized_2008}. The main technical difficulty is the construction of such a wave function, which is Gaussian--like far from the borders of the billiard and at the same time satisfies the Dirichlet boundary conditions, i.e. is zero at the border. Also, as one expects from physically viable wave function, it should be continuous and normalizable.

One of the options is to construct the coherent state as the superposition of the eigenstates in the billiard, an important example of such coherent states are the Klauder coherent states \cite{gazeau_coherent_1999, maioli_direct_2024}. The expansion coefficients can be chosen in various ways, e.g. with a Gaussian weighting \cite{fox_generalized_2000}. Construction of these states requires explicit calculation of the eigenfunctions and eigenenergies of the particle typically at large $n$. Also, crucially, the superposition is not guaranteed to be localized in a Gaussian wave packet style.

On the other hand, for some simple geometries the localized coherent state can be built using the method of images \cite{andrews_wave_1998, robinett_quantum_2004}. The most prominent examples are one--dimensional problems of a particle hitting a wall and confined in an infinite potential well. In this case, by adding and subtracting Gaussian wave packets in the exterior of the billiard, it is possible to build a localized wave function, which is zero on the boundary. This allows to avoid computation of the eigenstates and eigenergies entirely. The main disadvantage of the method is that, as mentioned, it is practically viable only for simple geometries, otherwise the structure of fictitious images becomes intractable for any reasonable calculation.

Our proposition is to define the coherent states in a way, that resembles both eigenfunction expansion and the method of images. In particular, we expand the Gaussian wave packet in the basis of eigenfunctions, continued to the exterior of the billiard, i.e. defined on the entire $\mathbb{R}^2$ and not only in the interior $\Omega$ of the billiard. The continuation, in turn, is performed by switching from differential Schroedinger equation to the equivalent Balian--Bloch integral equation  \cite{balian_distribution_1970, li_boundary_1995, tiago_boundary_1997, backer_behaviour_2002}. The solutions of the Balian--Bloch equation are well defined in the exterior of the billiard in contrast to the solutions of the Schroedinger equation, which are, strictly speaking, defined only in the interior. Also the Balian--Bloch equation can be understood as the generalization of the method of images to arbitrary boundaries.

We begin the paper with discussion of the continuation of the eigenfunctions of the particle in a quantum billiard by switching from the Schroedinger equation to the boundary integral Balian--Bloch equation. Next the coherent states, defined as the expansion of the minimal dispersion Gaussian wave packet in the basis of these continued eigenfunctions, are introduced. The construction of the coherent states is presented as the projection of a Gaussian to the subspace of the continued eigenstates. After that we use the connection between the Balian--Bloch equation and the Schroedinger equation following from the Hamiltonian with the surface delta--prime potential energy term. This allows to  construct the resolvent of the Balian--Bloch equation and propose a numerical procedure for computing the dynamics of the coherent states. Finally we consider the analytically solvable cases of the one--dimensional infinite well and Coxeter billiards, in which the coherent state wave functions are explicitly expressed via the Jacobi and Riemann theta functions.

\section{Continuation of the Schroedinger equation eigenfunctions}
The particle in the quantum billiard with the interior $\Omega \subset \mathbb{R}^2$ and the boundary $\partial \Omega$ is described by the free particle Hamiltonian together with the Dirichlet boundary conditions:
\begin{equation}
\begin{aligned}
&H_0 = \frac{p^2}{2m}\\
&\varphi(r) = 0, r \in \partial \Omega.
\end{aligned}
\end{equation}
Thus the corresponding time--independent Schroedinger equation is the Helmholtz equation 
\begin{equation}
\begin{aligned}
-&\frac{1}{2m} \partial^2_r \varphi_n(r) = E_n  \varphi_n(r)\\
&\varphi_n(r) = 0, r \in \partial \Omega.
\end{aligned}
\end{equation}
Strictly speaking, the eigenfunctions $\varphi_n(r)$ are defined only in the interior of the billiard, i.e. in the region $\Omega$. The continuation to the outside exists, but is not unique --- any other function defined  in the exterior, which satisfies the Helmholtz equation and matches the boundary condition, will be suitable. The usual choice in physics is setting the eigenfunctions outside of the billiard to be zero, i.e. represent the eigenfunctions as
\begin{equation}\label{eq:phi_ind}
\varphi_n(r) = \psi_n(r) \mathbb{1}_\Omega(r) = 
\begin{cases}
\psi_n(r), & r \in \Omega\\
0, & r \notin \Omega.
\end{cases}
\end{equation}
Here $\psi_n(r)$ is the function, satisfying the Helmholtz equation and the boundary conditions, $\mathbb{1}_\Omega(r)$  is the indicator function of the billiard.

Another choice is to build the continuation by also matching the normal derivative at the boundary. If one had obtained analytical expression for the eigenfunction in  form (\ref{eq:phi_ind}), then this continuation technically means removing the indicator function and leaving the function $\psi_n(r)$ as the solution. This is, however, not a satisfactory approach both from formal and practical points of view: ``removing the indicator'' is not a formal mathematical procedure and in most cases one can not find $\psi_n(r)$ analytically. Instead, the formal continuation procedure is switching to the equivalent integral equation with solutions, defined on the entire $\mathbb{R}^2$. For the time dependent Schroedinger equation in the billiard this is the Balian--Bloch integral equation \cite{balian_distribution_1970}
\begin{equation}\label{eq:BB}
\begin{aligned}
&\psi(r, t) = \psi_0(r, t) - \frac{1}{2m} \int\limits_0^t dt' \oint\limits_{\partial \Omega} \mathcal{G}_0(r - r', t - t') \partial_{n(r')} \psi(r', t') d \lambda_{r'}\\
&\psi_0(r, t) = \int\limits_{\mathbb R^2} \mathcal{G}_0(r - r', t) \psi(r', 0) d \lambda_{r'}\\
&\mathcal{G}_0(r, t) =\left(\frac{m}{2 \pi i t} \right)^\frac{D}{2} \exp \left\{\frac{i m r^2}{2 t}\right\}.
\end{aligned}
\end{equation}
Here $\psi(r, 0)$ is the initial wave function, $\mathcal{G}_0(r, t)$ is the free particle propagator in $D$ dimensions and $\partial_{n(r')} \psi(r', t')$ is the normal derivative at the boundary of the billiard, $d \lambda_{r'}$ is the line element and the index denotes the variable of integration.

\section{Coherent states in quantum billiards}\label{sec:Coh_st}
We propose the new construction of the coherent states $\Psi_c(r, t)$ in quantum billiards as the expansion of the minimal dispersion Gaussian wave packet $\Phi(r)$ in the basis of continued eigenfunctions $\psi_n(r)$, i.e.
\begin{equation}\label{eq:Psi_c_def}
\begin{aligned}
&\Psi_c(r, 0) = \sum_{n = 0}^\infty c_n \psi_n(r)\\
&c_n = \int\limits_{\mathbb R^2} \psi_n(r) \Phi(r) dr\\
&\Phi(r) = \frac{1}{\sqrt{2 \pi} \sigma} \exp \left\{-\frac{(r - r_0)^2}{4 \sigma^2} + i p_0 r \right\}.
\end{aligned}
\end{equation}
This means that the time dependence of the coherent state is governed by the Balian--Bloch equation (\ref{eq:BB}) with the initial wave function $\Phi(r)$:
\begin{equation}
\Psi_c(r, t) = \int\limits_{\mathbb R^2} \mathcal{G}_0 (r - r')\Phi(r') d \lambda_{r'}- \frac{1}{2m} \int\limits_0^t dt' \oint\limits_{\partial \Omega} \mathcal{G}_0(r - r', t - t') \partial_{n(r')} \Psi_c(r', t') d \lambda_{r'}.
\end{equation}
This particular construction is motivated, as discussed in the Introduction, by its connection both to the method of images and to the eigenstate expansion. Namely, the Balian--Bloch equation can be understood as the generalization of the method of images. Another motivation is the continuity of the coherent state wave function at the boundary. In particular, let us consider the expansion of the Gaussian wave packet $\Phi(r)$ in (\ref{eq:Psi_c_def}) in the basis of eigenfunctions (\ref{eq:phi_ind}), which are zero outside of the billiard:
\begin{equation}\label{eq:Theta}
\Xi(r) = \sum_{n = 0}^\infty \varphi_n(r) \int\limits_{\mathbb R^2} \varphi_n(r') \Phi(r') dr' =  \sum_{n = 0}^\infty \psi_n(r) \mathbb{1}_\Omega(r) \int\limits_{\mathbb R^2} \psi_n(r') \Phi(r') \mathbb{1}_\Omega(r') dr'.
\end{equation}
The wave functions $\varphi_n(r)$ form a complete basis inside $\Omega$, so
\begin{equation}
\sum_{n = 0}^\infty \varphi_n(r) \varphi_n(r') \Big|_{r,r' \in \Omega} = \delta(r - r') \Big|_{r,r' \in \Omega} \Rightarrow \Theta(r) = \int\limits_{\mathbb R^2} \delta(r - r') \Phi(r') \mathbb{1}_\Omega(r') dr' = \Phi(r) \mathbb{1}_\Omega(r).
\end{equation}
Given that $\Phi(r \in \Omega) \neq 0$, the function $\Xi(r)$ has a jump discontinuity at the boundary of the billiard. Thus, if we define the coherent state in the billiard as the expansion of the Gaussian, the continuity condition makes it impossible to define the integration domain of the expansion coefficients as the interior of the billiard. Definition (\ref{eq:Psi_c_def}), i.e. expansion in the basis of continued eigenfunctions, avoids this problem. One might also note the singular behaviour of the function (\ref{eq:Theta}): all partial sums are zero for $r \in \partial \Omega$ due to $\varphi_n(r \in \Omega) = 0$, however, the total sum with infinite number of terms is non--zero. This a phenomenon, common in the case of expansion of a function in the basis, defined only in some bounded region. The most prominent example is the Fourier series, which we will encounter, when considering the particle in a one--dimensional infinite potential well, see sec. \ref{sec:Well} and fig. \ref{fig:disc_well}. The discontinuity can be avoided, if the summation over the eigenstates is restricted to some finite upper limit $N$, then the result will be zero at the boundary due to summation of a finite number of zeros. The price is the Gibbs phenomenon --- rapid oscillations of the result near the boundaries. In addition, if one is interested in classical limit of the motion in the billiard, it implies the high energy limit and consequently $n \rightarrow \infty$, thus summation over all eigenstates is desired.

\subsection{Coherent states as projections}\label{sec:proj}
These coherent states can be thought of as the projections of the minimal dispersion wave packet on the subspace, formed by the continued eigenstates of the particle in the billiard. In particular,
\begin{equation}\label{eq:P}
\begin{aligned}
&\Psi_c(r, 0) = P \Phi(r)\\
&P = \sum_n |\psi_n \rangle  \langle \psi_n|.
\end{aligned}
\end{equation}
Looking at the expression for the projection operator $P$, one might think that it is just an identity operator due to completeness of the eigenfunction basis $\{\psi_n(r) \}$. However, this holds only for functions in the function space $\operatorname{span}\{\psi_n(r) \}$, i.e.  the  linear combinations of $\{\psi_n(r) \}$. For a general function from $\mathbb{L}^2$ the action of $P$ is not trivial.

The operator $P$ can be expressed via the evolution operator of the particle
\begin{equation}
G = \sum_n e^{-i E_n t} |\psi_n \rangle  \langle \psi_n|,
\end{equation}
so it follows that
\begin{equation}\label{eq:P_lim}
P = \lim\limits_{t \rightarrow 0^+} G.
\end{equation}
The kernel of the operator $G$ can be also obtained from the Balian--Bloach equation (\ref{eq:BB}) if one substitutes $\Psi_c(r, t) = \int \mathcal{G}(r, r', t) \Phi(r') dr'$. This gives
\begin{equation}
\mathcal{G}(r, r', t) = \mathcal{G}_0(r, r', t)  - \frac{1}{2m}  \int\limits_0^t  dt' \oint\limits_{\partial \Omega} \mathcal{G}_0(r, r', t - t') \partial_{n(r')} \mathcal{G}(r', r'', t')  d \lambda_{r''}.
\end{equation}
Notably, $\mathcal{G}(r, r', 0) \neq \delta(r - r')$, i.e. the evolution operator is not identity at $t = 0$, because the Gaussian $\Phi(r)$ does not satisfy the Dirichlet boundary conditions and can not be the allowed state in the quantum billiard. Instead it gets immediately projected on the basis of the continued eigenstates $\psi_n(r)$, which is reflected by taking the limit at $0^+$ in (\ref{eq:P_lim}). Formally this can be seen if we consider the first term of the Dyson series:
\begin{equation}
\begin{aligned}
&\mathcal{G}(r, r', t) \approx \mathcal{G}_0(r, r', t)  - \frac{1}{2m}  \int\limits_0^t  dt' \oint\limits_{\partial \Omega} \mathcal{G}_0(r, r', t - t') \partial_{n(r')} \mathcal{G}_0(r', r'', t')  d \lambda_{r''} = \\
&= \mathcal{G}_0(r, r', t) + \frac{i}{2 m^2}  \int\limits_0^t  dt' \oint\limits_{\partial \Omega} \frac{1}{t'} \mathcal{G}_0(r, r', t - t') \mathcal{G}_0(r', r'', t')(r' - r'') n(r'') d \lambda_{r''}.
\end{aligned}
\end{equation}
We have used the definition  of the normal derivative $\partial_{n(r')} \mathcal{G}_0(r', r'', t') = \nabla_{r'} \mathcal{G}_0(r', r'', t') n(r'')$, where $n(r'')$ is the normal to the boundary $\partial \Omega$ at point $\{x'', y''\}$. Now in the limit of $t \rightarrow 0^+$ we find
\begin{equation}
\begin{aligned}
&\mathcal{G}(r, r', 0) \approx \mathcal{G}_0(r, r', 0) - \frac{i}{2 m} \lim\limits_{t \rightarrow 0^+} t \frac{\partial}{\partial t} \int\limits_0^t I(r, r', t, t') d t'\\
&I(r, r', t, t') = \frac{1}{t'} \oint\limits_{\partial \Omega} \mathcal{G}_0(r, r', t - t') \mathcal{G}_0(r', r'', t') (r' - r'') n(r'') d \lambda_{r''}.
\end{aligned}
\end{equation}
The first term gives the delta--function kernel: $\mathcal{G}_0(r, r', 0) = \delta(r - r')$. The second term adds a non--zero contribution even in the limit of zero upper limit of the integral, because the integrand is singular. In particular, 
\begin{equation}
\lim\limits_{t \rightarrow 0^+} t \frac{\partial}{\partial t} \int\limits_0^t I(r, r', t, t') = \lim\limits_{t \rightarrow 0^+} t \left[ I(r, r', t, t) + \int\limits_0^t \partial_t I(r, r', t, t') d t'\right].
\end{equation}
The first term is $\propto \oint \mathcal{G}_0(r, r', 0) \mathcal{G}_0(r', r'', 0) \propto \oint \delta(r - r') \delta (r' - r'')$, so this is a distribution, supported on the boundary $\partial \Omega$. The second term can be evaluated in the same way by expanding the integral in the Taylor series.

\subsection{A note on normalization}
The coherent state (\ref{eq:Psi_c_def}) should also be normalized with integration restricted to the interior of the billiard. Indeed, let us compute the norm:
\begin{equation}
|\!| \Psi_c(r, 0) |\!| = \sum_{nm} c_n^* c_m \int\limits_\Omega \psi_n^*(r) \psi_m(r) dr = \sum_{nm} c_n^* c_m \delta_{nm} = \sum_n |c_n|^2.
\end{equation}
The integral was collapsed to the delta symbol due to orthogonality of the wave functions $\psi_n(r)$ in the interior of the billiard. In general, the resulting sum is not equal to unity, so one has to additionally multiply the coherent state wave function by the normalization coefficient
\begin{equation}
A = \left( \sum_n |c_n|^2 \right)^{-1/2}.
\end{equation}
In the manuscript we suppress the normalization coefficient for sake of brevity and refer to the non--normalized wave function (\ref{eq:Psi_c_def}) as the coherent state wave function. Crucially, the normalization coefficient does not depend on time, because for any time $t > 0$ the coherent state $\Psi_c(r, t) \in \operatorname{span}\{\psi_n(r) \}$ and the quantum evolution in the billiard for the wave functions from this functional space is unitary.

\subsection{So, what is the wave function outside of the billiard?}
We have defined the coherent states that are non--zero outside of the billiard and one might wonder does this make a difference from a physical point of view, compared to the wave function, which is cut off outside the billiard \cite{wheeler_2000}. The answer depends on the definition of physical observables, namely the definition of the limits of the corresponding integral. In any case, it is only physically correct to perform the integration over the exterior of the billiard, which can be formally written in two ways:
\begin{equation}
\langle A \rangle = \int\limits_\Omega \psi^*(r) A \psi(r) dr \text{ and } \langle A \rangle = \int\limits_{\mathbb{R}^2} \psi^*(r) A \psi(r) \mathbb{1}_\Omega dr.
\end{equation}
This means, that multiplication by the indicator function $\mathbb{1}_\Omega$, i.e. cutting off outside, is necessary if the average is defined as an integral over the whole $\mathbb{R}^2$. Otherwise, if the integration is restricted to $r \in \Omega$, the values of the wave function outside have no effect on physical observables. This, however, comes with the cost of poorly defined momentum operator, the standard definition $p = -i \partial_x$ of which becomes non self--adjoint \cite{vincenzo_2008, al-hashimi_canonical_2021}.

A peculiarity also appears when defining the Wigner function of the particle in a quantum billiard. Its formal definition implies integration over the whole $\mathbb{R}^2$, so if one sticks to it, it is required to introduce the multiplication by the indicator function $\mathbb{1}_\Omega$. The latter leads to imposing the boundary conditions procedure using the convolution \cite{seidov_wigner_2023, seidov_wigner_2025, burkov_features_2025}.

In the present manuscript the coherent state wave functions are continued outside the billiard, so the averages should be defined as integrals, restricted to the billiard. However, it might be beneficial to switch to an alternative approach and multiply the wave functions by the indicator function, for example, when carrying out calculations in the phase--space quantization formalism.

\section{Practical construction of coherent states and their dynamics}
The construction of the coherent states then boils down to solution of the Balian--Bloch equation (\ref{eq:BB}) with the Gaussian initial conditions. This can be done by various numerical and in some simple cases analytical methods. We propose an approach which exploits the fact that the equation (\ref{eq:BB}) can be derived from the Hamiltonian 
\begin{equation}\label{eq:H_delta}
H = \frac{p^2}{2 m} + \frac{1}{2 m}\delta'_{\partial \Omega}(r).
\end{equation}
This was shown in \cite{lange_potential_2012, lange_distribution_2015} and the derivation with different approach is presented in the Appendix. Here $\delta'_{\partial \Omega}(r)$ is the derivative of the surface delta--function, supported on the boundary of the billiard. Formally it is defined as $\delta'_{\partial \Omega}(r) = \nabla^2 \mathbb{1}_\Omega$, where $\mathbb{1}_\Omega$ is the indicator function of the billiard. It has the property
\begin{equation}\label{eq:delta_s}
\int \delta'_{\partial \Omega}(r) f(r) dr = - \oint\limits_{\partial \Omega} \partial_n f(r) d \lambda_{r}.
\end{equation}
In the one--dimensional case the equivalence between the motion in the billiard and the $\delta'_{\partial \Omega}(r)$  potential was shown in \citep{dias_wigner_2002, dias_boundaries_2021}, the multidimensional case was considered in \cite{seidov_wigner_2025}.

So, the dynamics of the coherent state is governed by usual evolution operator equation, where the initial state is the minimal dispersion Gaussian wave packet $\Phi(r)$:
\begin{equation}
\begin{aligned}
&\Psi_c(r, t) = U(t) \Phi(r)\\
&U(t) = \exp\left\{-i H t \right\}  = \exp\left\{-\frac{i t}{2m} \left[p^2 + \delta'_{\partial \Omega} (r) \right] \right\}.
\end{aligned}
\end{equation}
This is a formal expression which is troublesome to interpret, given that the exponential is taken from a singular object. However, the procedure for calculating $\Psi_c(r, t)$ can be defined. First, we perform the Laplace transform with respect to time, this gives
\begin{equation}
\mathcal{L}_{t \rightarrow s} \Psi_c(r, t) = \tilde \Psi_c(r, s) = \frac{2 i m}{2 i m s - p^2 - \delta'_{\partial \Omega}(r)} \Phi(r).
\end{equation}
Next with the Fourier transform with respect to the spatial coordinate $r$ we find
\begin{equation}\label{eq:Psi_c_ps}
\mathcal{F}_{r \rightarrow p} \tilde \Psi_c(r, s) = \hat{\tilde{\Psi}}_c(p, s) = \frac{2 i m}{2 i m s - p^2 - \Lambda} \hat \Phi(p).
\end{equation}
The fraction is in fact the resolvent of the Balian--Bloch equation. Here $\Lambda$ is the convolution operator, defined as convolution with the Fourier transform of $\delta'_{\partial \Omega}(r)$. In particular,
\begin{equation}
\begin{aligned}
&\Lambda f(p) = \hat \delta'_{\partial \Omega}(p) *_p f(p) = \int \hat \delta'_{\partial \Omega}(p - p') f(p') dp'\\
&\hat \delta'_{\partial \Omega}(p) = - \oint\limits_{\partial \Omega} \partial_n e^{i p r} d \lambda_r.
&\end{aligned}
\end{equation}
This Fourier image of the derivative of the surface delta--function is not singular and is well behaved. Together this defines a well behaved image of the coherent state wave function $\hat{\tilde{\Psi}}_c(p, s)$ in the $\{p, s\}$ space. The coherent state wave function itself is then found by performing the inverse transforms, i.e.
\begin{equation}
\Psi_c(r, t) = \mathcal{L}^{-1}_{s \rightarrow t} \mathcal{F}^{-1}_{p \rightarrow r} \hat{\tilde{\Psi}}_c(p, s) =  \mathcal{L}^{-1}_{s \rightarrow t} \mathcal{F}^{-1}_{p \rightarrow r} \left\{\frac{2 i m}{2 i m s - p^2 - \Lambda} \hat \Phi(p) \right\}.
\end{equation}

\subsection{Numerical implementation}
The main computational problem is to implement the action of the resolvent operator
\begin{equation}
R =  \frac{2 i m}{2 i m s - p^2 - \Lambda} 
\end{equation}
on the initial Gaussian wave function in (\ref{eq:Psi_c_ps}). We have chosen the discretization approach in which the operators are presented as matrices. The matrix elements are values of the operator kernels at grid points $p_x^j$, $p_y^j$, ${p'_x}^j$, ${p'_y}^j$, the index $j$ takes values at $N$ points such that $p_{x,y} \in [-P, P]$. The wave functions are discretized accordingly and are represented as vectors. So, the resolvent is represented by a matrix
\begin{equation}
\mathbf{R} = 2 i m \left(2 i m s \mathbf{1} - \mathbf p^2 - \mathbf \Lambda \right)^{-1}.
\end{equation}
The bold font denotes the matrices of the linear operators. This leads to the coherent state wave function vector
\begin{equation}
\hat{\tilde{\mathbf{\Psi}}}_c = \mathbf R \hat{\mathbf{\Phi}}.
\end{equation}

The transformation to the real space from the momentum space is performed using the FFT (fast Fourier transform) algorithm. This is very efficient, but does not allow to find values of $\tilde \Psi_c(r, s)$ at the border, when $r \in \partial \Omega$. To obtain the latter, direct numerical integration is required. However, given that $\tilde \Psi_c(r \in \partial \Omega, s) = 0$, the numerical integration converges poorly, still one might wish to carry it out in order to estimate the numerical error by deviation of the numerical value from zero.

Next, the time dependence should be found by performing numerically the inverse Laplace transform. However, instead we compute the short term behaviour using the Laplace initial value theorem. According to it
\begin{equation}
\Psi_c(r, 0^+) = \lim\limits_{s \rightarrow \infty} s \tilde \Psi_c(r, s).
\end{equation}
Also this defines the projection operator in (\ref{eq:P}) as
\begin{equation}
P = \lim\limits_{s \rightarrow \infty} s R.
\end{equation}
Numerically, the limit is understood as choosing the value $s \gg |\!|\mathbf p^2 + \mathbf \Lambda|\!|$, where $|\!|\cdot |\!|$ is the norm of the matrix --- its largest by absolute value singular value.

\subsection{Numerical example of coherent states}
\begin{figure}[h!!]
\includegraphics[width=0.49\textwidth]{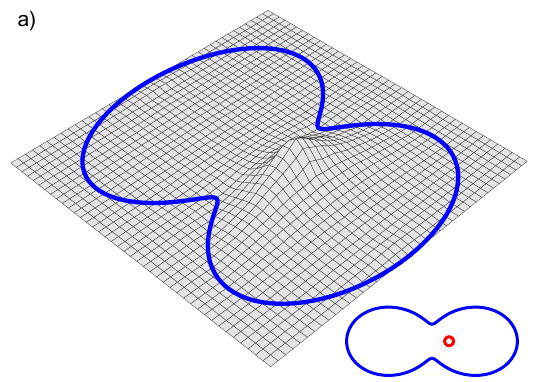}
\includegraphics[width=0.49\textwidth]{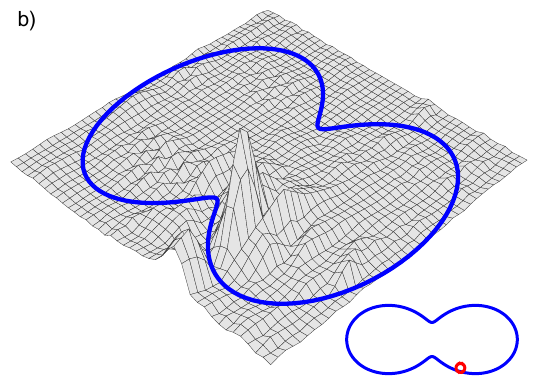}
\caption{The absolute values of the coherent states wave functions in the billiard with boundary given by $r(\theta) = \cos^2 \theta + (1/5) \sin^4 \theta$ in polar coordinates (blue curve). In the insets the initial Gaussian wave packet (red circle) is depicted relative to the boundary. The diameter of the circle corresponds to one standard deviation of the absolute value of the initial Gaussian wave packet. In a) the coherent state is placed away from the boundary, so the wave function resembles a Gaussian. In b) the coherent state is placed near the boundary, so the oscillations of the wave function are present.}
\label{fig:coh_st_num}
\end{figure}

In fig. \ref{fig:coh_st_num} we present the numerically obtained coherent states at $t = 0$ in the billiard with the boundary, defined by equation $r(\theta) = \cos^2 \theta + (1/5) \sin^4 \theta$ in polar coordinates. In fig. \ref{fig:coh_st_num}a the coherent state is placed at $x_0 = 1/5$ and $y_0 = 0$. This is fairly far from the boundary (see the inset), so the wave function resembles the free Gaussian. In contrast, in fig. \ref{fig:coh_st_num}b the coherent state is placed near the boundary at $x_0 = 1/3$, $y_0 = -1/3$ and the coherent state wave function has developed oscillations, which is expected for the state near the boundary. One can also observe non--zero values of the wave functions outside of the billiard --- the consequence of the expansion in the continued eigenstates basis. In principle, nothing is preventing us from placing the initial Gaussian wave packet outside of the billiard, the equation (\ref{eq:BB}) will still have a solution. This will describe the quantum particle, scattering from an obstacle in a shape of the billiard. A scattering problem using boundary integral methods was considered in \cite{maioli_exact_2018}, but using the surface $\delta$--potential instead of the surface $\delta'$--potential.

\section{Analytical examples}
\subsection{One--dimensional infinite potential well}
Let us first consider the problem of a one--dimensional potential well, i.e. a particle confined to the interval $x \in [0, L]$. The continued eigenfunctions and the one--dimensional Gaussian are respectively
\begin{equation}
\begin{aligned}
&\psi_n(x) = \sqrt\frac{2}{L} \sin\left(\frac{\pi n x}{L} \right), \ \ x \in \mathbb{R}\\
&\Phi(x) = \frac{1}{\sqrt{\sqrt{2 \pi} \sigma}} e^{-\frac{(x - x_0)^2}{4 \sigma^2} + i p_0 x}.
\end{aligned}
\end{equation}
Given that the eigenenergies of the particle in a one--dimensional potential well are also known, we can write the coherent states as
\begin{equation}
\begin{aligned}
&\Psi_c(x, t) = \sum_{n = 0}^\infty c_n e^{-i E_n t} \psi_n(x)\\
&E_n = \frac{\pi^2 n^2}{2 m L^2}\\
&c_n = \int\limits_\mathbb{R} \Phi(x') \psi_n(x') dx'.
\end{aligned}
\end{equation}
It is convenient to carry out the summation first, i.e. rearrange the expression as
\begin{equation}\label{eq:Psi_c_well}
\Psi_c(x, t) = \frac{2}{L}\int\limits_\mathbb{R} dx' \Phi(x') \sum_{n=0}^\infty e^{-\frac{i \pi^2 n^2}{2 m L^2} t} \sin \left(\frac{\pi n x}{L} \right)\sin \left(\frac{\pi n x'}{L} \right).
\end{equation}
In fact, the sum is the propagator of the particle in the well:
\begin{equation}\label{eq:G_well}
\begin{aligned}
\mathcal G_{A_1}(x, x', t) &= \frac{2}{L}\sum_{n=0}^\infty e^{-\frac{i \pi^2 n^2}{2 m L^2} t} \sin \left(\frac{\pi n x}{L} \right)\sin \left(\frac{\pi n x'}{L} \right) =\\
& = \frac{1}{L} \sum_{n=0}^\infty e^{-\frac{i \pi^2 n^2}{2 m L^2} t} \cos \left(\frac{\pi n}{L}[x - x'] \right) - \frac{1}{L} \sum_{n=0}^\infty e^{-\frac{i \pi^2 n^2}{2 m L^2} t} \cos \left(\frac{\pi n}{L}[x + x'] \right) = \\
& = \frac{1}{2L} \theta_3 \left(\frac{\pi}{2L} (x - x'), e^{-\frac{i \pi^2 t}{2 L^2 m}} \right) - \frac{1}{2L} \theta_3 \left(\frac{\pi}{2L} (x + x'), e^{-\frac{i \pi^2 t}{2 L^2 m}} \right).
\end{aligned}
\end{equation}
Here $\theta_3(z, q)$ is the Jacobi theta function. The subscript $A_1$ is related to the Coxeter group $A_1$, which will be relevant further, when we will consider $2d$ Coxeter billiards. This expression can be also obtained using the method of images, which was done in \cite{gori_propagator_2001}. Notably, as discussed in section \ref{sec:proj}, this propagator is not just the $\delta(x - x')$ at $t \rightarrow 0$, instead it is the difference of the Dirac combs. This can be seen by setting $t = 0$ in the sum:
\begin{equation}
\begin{aligned}
\mathcal G_{A_1}(x, x', 0) &= \frac{2}{L}\sum_{n=0}^\infty\sin \left(\frac{\pi n x}{L} \right)\sin \left(\frac{\pi n x'}{L} \right) =\\
&= \frac{1}{2L} \sum_{n=0}^\infty \left\{e^{-\frac{i \pi n}{L}(x - x')} + e^{\frac{i \pi n}{L}(x - x')} - e^{-\frac{i \pi n}{L}(x + x')} - e^{\frac{i \pi n}{L}(x + x')}\right\} = \\
& = \frac{1}{2L} \sum_{n = -\infty}^\infty \left\{e^{-\frac{i \pi n}{L}(x - x')} -  e^{\frac{i \pi n}{L}(x + x')}\right\}.
\end{aligned}
\end{equation}
Next by using the identity
\begin{equation}
\frac{1}{2L} \sum_{n = -\infty}^\infty e^{\frac{2 i \pi n x}{L}} = \frac{1}{2L} \sum_{n = -\infty}^\infty \delta \left(\frac{x}{2L} - n \right) = \Sha_{2L}(x)
\end{equation}
we find
\begin{equation}
\mathcal G(x,  x', 0) = \Sha_{2L}(x - x') - \Sha_{2L}(x + x').
\end{equation}
It is just $\delta(x - x')$ if restricted to the interior of the well, in which case the delta functions outside do not contribute.

\begin{figure}[h!!]
\center\includegraphics[width = \textwidth]{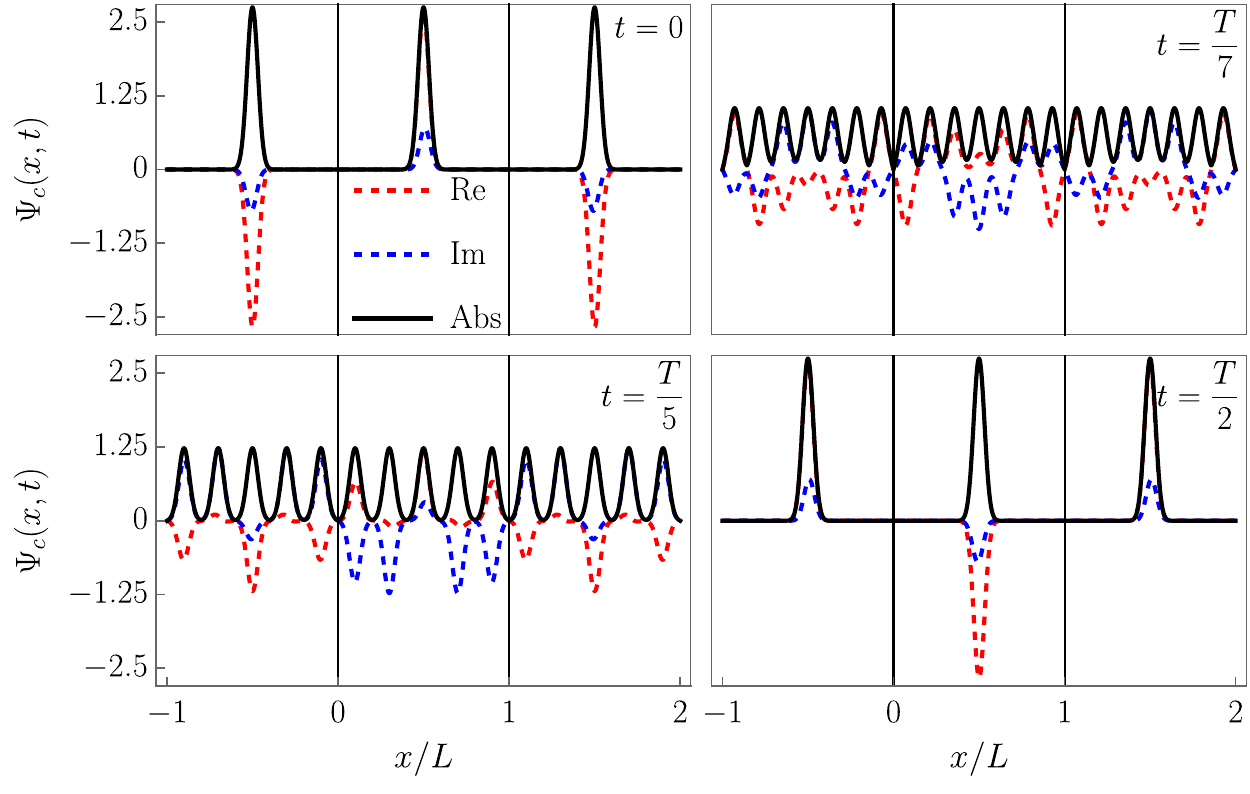}
\caption{Coherent state (\ref{eq:coh_st_well}) of the particle in a one--dimensional infinite potential well at different times. The time is expressed in the units of the period of motion $T= 4 m L^2/\pi$, the initial wave packet dispersion is $\sigma = 0.045L$, the vertical lines represent the walls of the well. The plots are made in the extended picture, i.e the wave function is not cut off outside of the well, so one can  observe the continuation to the whole $\mathbb{R}$.}
\label{fig:coh_st_well}
\end{figure}

Now the integral has to be calculated, which in essence is the convolution of the Gaussian $\Phi(x)$ with the Jacobi theta function and thus equals to another theta function. We use the series representation of the theta--function:
\begin{equation}
\begin{aligned}
&\int\limits_\mathbb{R} \theta_3\left(\frac{\pi (x \pm x')}{2L}, q \right) e^{-\frac{(x' - x_0)^2}{4 \sigma^2} + i p_0 x'} dx' = \sum_{n = -\infty}^\infty q^{n^2} \int\limits_\mathbb{R}  e^{\frac{i \pi n (x \pm x')}{L}} e^{-\frac{(x' - x_0)^2}{4 \sigma^2} + i p_0 x'} dx' =\\
&= 2 \sigma \sqrt\pi e^{i p_0 x_0 - p_0^2 \sigma^2}\sum_{n = -\infty}^\infty q^{n^2} \exp\left\{- \frac{\pi^2 n^2 \sigma^2}{L^2} + \frac{i \pi n}{L} \left(x \pm x_0 \pm 2i p_0 \sigma^2 \right) \right\} =\\
&= 2 \sigma \sqrt\pi e^{i p_0 x_0 - p_0^2 \sigma^2} \theta_3 \left(\pi \frac{x \pm x_0 \pm 2 i p_0 \sigma^2}{2L}, e^{-\frac{\sigma^2 \pi^2}{L^2}} q \right).
\end{aligned}
\end{equation}
Finally this gives the coherent state of the particle in a one--dimensional infinite potential well in form
\begin{equation}\label{eq:coh_st_well}
\begin{aligned}
&\begin{aligned}
\Psi_c(x, t) &= \frac{1}{L} \sqrt{\sigma \sqrt\frac{\pi}{2}} e^{i p_0 x_0 - p_0^2 \sigma^2} \times\\
&\times \left\{\theta_3 \left(\pi \frac{x - x_0 - 2 i p_0 \sigma^2}{2L},q' \right) - \theta_3 \left(\pi \frac{x + x_0 + 2 i p_0 \sigma^2}{2L},  q' \right) \right\}
\end{aligned}\\
&q' = \exp\left\{-\frac{\sigma^2 \pi^2}{L^2} -\frac{i \pi^2 t}{2 L^2 m} \right\}. 
\end{aligned}
\end{equation}
The plots of this function are presented in fig.\ref{fig:coh_st_well}. One can observe, that the initially localized wave packet delocalizes. The motion is periodic with the period $T = 4 m L^2/\pi$. Interesting enough, this type of coherent states were considered in \cite{doncheski_wave_2003}, but the authors treated the integration over entire $\mathbb{R}$ as an approximation. Here we claim that this is not an approximation, but indeed the valid definition of coherent states. 

\subsubsection{Expansion in the bounded eigenfunctions basis}\label{sec:Well}
\begin{figure}[h!!]
\center\includegraphics[width = 0.49\textwidth]{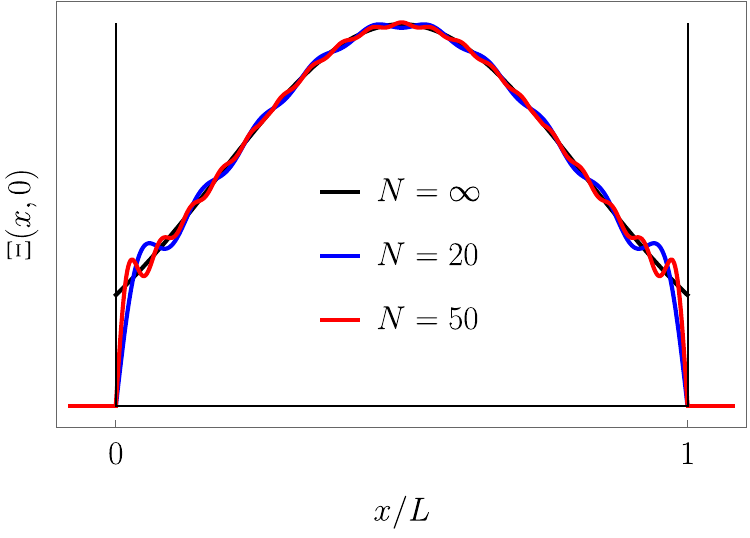}
\caption{The function $\Xi(x, 0)$ and partial sums with $N = 20$ and $N = 50$ terms, eq. (\ref{eq:Theta_well}). One can see, that if the sum is performed in infinite limits, the wave function is discontinuous at the boundaries of the well.}
\label{fig:disc_well}
\end{figure}

Let us illustrate the discontinuity problem, which arises if one limits the integration to the interior of the well, as discussed for the general case in sec. \ref{sec:Coh_st}. So, we define the function
\begin{equation}\label{eq:Theta_well}
\Xi(x, t) = \frac{2}{L}\int\limits_\mathbb{R} dx' \Phi(x') \sum_{n=0}^\infty e^{-\frac{i \pi^2 n^2}{2 m L^2} t} \sin \left(\frac{\pi n x}{L} \right)\sin \left(\frac{\pi n x'}{L} \right) \mathbb{1}_\text{well}(x) \mathbb{1}_\text{well}(x').
\end{equation}
With calculations, completely analogous to the ones carried out in the previous section, we find
\begin{equation}
\Xi(x, 0) = \mathbb{1}_\text{well}(x) \frac{2}{L}\int\limits_0^L dx' \Phi(x') \Big\{ \Sha_{2L}(x - x') - \Sha_{2L}(x + x') \Big\} dx' = \Phi(x) \mathbb{1}_\text{well}(x)
\end{equation}
It is clearly discontinuous at $x = \pm L$, as one can observe on the plot in fig. \ref{fig:disc_well}. The plot is accompanied with the plots of the function $\Theta(x, 0)$ but with finite number $N$ of terms in the sum. The latter are continuous and equal to zero at the boundary.

\subsection{Coxeter billiards}
A less trivial, but still analytically solvable cases are the so--called Coxeter billiards. These are the billiards in shape of figures, suitable for tiling the $2d$ plane. We can find the corresponding propagators through the method of images for a several fundamental Coxeter billiards: a rectangle (Coxeter group $A_1 \times A_1$), a $45$--$45$--$90$ triangle (Coxeter group $B_2$), a $60$--$60$--$60$ triangle (Coxeter group $A_2$) and a $30$--$60$--$90$ triangle (Coxeter group $G_2$).
\subsubsection{Rectangular well}
The rectangular well with dimensions $L_1$ and $L_2$ can be constructed as a product of two orthogonal one--dimensional wells, so the corresponding propagator is the product of one--dimensional well propagators (\ref{eq:G_well}):
\begin{equation}\label{eq:prop_rect}
\begin{aligned}
\mathcal{G}_{A_1 \times A_1}\left(r, r', t\right) &= \mathcal{G}_{A_1}(x, x', t) \cdot \mathcal{G}_{A_1}(y, y', t) \\
&\propto \sum_{\omega \in W_{A_1 \times A_1}} \det(\omega) \ \Theta\left(\sigma_{A_1 \times A_1} \left(\omega r-r'\right), -\dfrac{t}{2 m} \sigma_{A_1 \times A_1}^2\right).
\end{aligned}
\end{equation}
Here 
\begin{equation}
\begin{aligned}
&\sigma_{A_1 \times A_1} = \dfrac{\pi}{2}\begin{pmatrix} 1/L_1  & 0\\ 0 & 1/L_2 \end{pmatrix}
\end{aligned}
\end{equation}
and $w \in W$ is the element of the matrix representation $W$ of the Weyl group, corresponding to the considered billiard. The function $\Theta(u, M)$ is the multivariate Riemann (or also called Siegel) theta function \cite{dubrovin_theta_1981,deconinck_computing_2002,wheeler1997appliedtheta}. The coherent state wave function now can be also represented as the Riemann theta function by performing the convolution of the propagator with a Gaussian wave packet, analogous to (\ref{eq:coh_st_well}) in the one--dimensional well.

\subsubsection{$45$--$45$--$90$ triangle well}
Now we can split the square well into two triangle wells along the diagonal. Then by applying the method of images with respect to the diagonal we obtain the propagator in a $45$--$45$--$90$ triangle well:
\begin{equation}\label{eq:prop_iso_right_tri}
\begin{aligned}
\mathcal{G}_{B_2}\left(r, r', t\right) &= \mathcal{G}_{A_1 \times A_1}\left(r, r', t\right) - \mathcal{G}_{A_1 \times A_1}\left(\mathrm{Ref}(\pi/4) r, r', t\right) \\
&\propto \sum_{\omega \in W_{B_2}} \det(\omega) \ \Theta\left(\sigma_{A_1 \times A_1} \left(\omega r-r'\right), -\dfrac{t}{2 m} \sigma_{A_1 \times A_1}^2 \right).
\end{aligned}
\end{equation}
Here
\begin{equation}
\begin{aligned}
\mathrm{Ref}(\alpha) = \begin{pmatrix} \cos(2\alpha)  & \sin(2\alpha)\\ \sin(2\alpha) & -\cos(2\alpha) \end{pmatrix}.
\end{aligned}
\end{equation}

\subsubsection{$60$--$60$--$60$ triangle well}
The equilateral triangle requires more tedious calculations, which can be found in \cite{wheeler1997particlebox}. The result gives the propagator
\begin{equation}
\begin{aligned}
\mathcal{G}_{A_2}\left(r, r', t\right) &\propto \sum_{\omega \in W_{A_2}} \det(\omega) \ \Theta\left(\sigma_{A_2} \left(\omega r - r'\right), -\dfrac{t}{2 m} \sigma_{A_2}^2 \right) +\\
&+ \sum_{\omega \in W_{A_2}} \det(\omega) \ \Theta\left(\sigma_{A_2} \left(\omega r - r' + v\right), -\dfrac{t}{2 m} \sigma_{A_2}^2 \right).
\end{aligned}
\end{equation}
Here 
\begin{equation}
\begin{aligned}
&\sigma_{A_2} = \dfrac{\pi}{3L}\begin{pmatrix} 1  & 0\\ 0 & \sqrt{3} \end{pmatrix},
&v = \dfrac{L}{2} \begin{pmatrix} 3  \\ \sqrt{3} \end{pmatrix}.
\end{aligned}
\end{equation}
\subsubsection{$30$--$60$--$90$ triangle well}
In the same way as the propagator in $45$--$45$--$90$ triangle well was found by splitting the square, we can split the $60$--$60$--$60$ triangle into two $30$--$60$--$90$ triangles and find the propagator
\begin{equation}
\begin{aligned}
\mathcal{G}_{G_2}\left(r, r', t\right) &= \mathcal{G}_{A_2}\left(r, r', t\right) - \mathcal{G}_{A_2}\left(\mathrm{Ref}(\pi/6)r, r', t\right) \\
&\propto \sum_{\omega \in W_{G_2}} \det(\omega) \ \Theta\left(\sigma_{A_2} \left(\omega r - r'\right), -\dfrac{t}{2 m} \sigma_{A_2}^2 \right) +\\
&+ \sum_{\omega \in W_{G_2}} \det(\omega) \ \Theta\left(\sigma_{A_2} \left(\omega r - r' + v\right), -\dfrac{t}{2 m} \sigma_{A_2}^2 \right).
\end{aligned}
\end{equation}

\subsubsection{Relation to the expansion in the basis of continued eigenfunctions}
These results, analogous to the case of the one--dimensional well, can be found by representing the eigenfunctions as superpositions of plane waves. In particular, it is known, that in the Coxeter billiards the eigenfunctions are given by \cite{turner_quantum_1984, terras_image_1980, li_particle_1987, wheeler1997particlebox}:
\begin{equation}
\psi_k(r) = \sum_{\omega \in W} \det(\omega) e^{i (\omega k) r}.
\end{equation}
Here the eigenfunctions domain is extended outside of the billiard, i.e. $r \in \mathbb R^2$. Next, if one substitutes these eigenfunctions in the expansion of the propagator, all the same results for the propagators will follow from simplifications of the sums and their reduction to the Riemann theta functions \cite{wheeler1997particlebox}.

\section{Conclusions} 
The generalized coherent states in quantum billiards, proposed in the present manuscript, are, in essence, projections of the Gaussian wave packet to the functional space, spanned by the continued eigenfunctions of the particle in the billiard. The continuation is defined by matching not only the value of the wave function itself at the boundary, but also its normal derivative. Formally such continuation can be built if one switches from the Schroedinger equation to the equivalent Balian--Bloch boundary integral equation. The generalized coherent state is then the solution of the Balian--Bloch equation with the Gaussian wave packet as the initial state. Projection to the basis of continued eigenstates avoids the singular behaviour of the expansion at the boundaries of the billiard.

We build the resolvent of the Balian--Bloch equation by exploiting its connection to the Schroedinger equation with the surface delta--prime potential energy term. This allows to develop the numerical procedure of construction of such coherent states in billiards of arbitrary shape. Moreover, the initial wave packet can be placed outside of the billiard, which gives the solution of the scattering problem. Also the expression for the resolvent allows to perform analytical investigations using standard approaches of the resolvent formalism.

Finally, the analytical examples of the one--dimensional infinite potential well and Coxeter billiards are considered. In this cases the wave functions of the coherent states can be expressed explicitly in terms of Jacobi theta functions. This is not a coincidence, because of the deep connection between the theta functions and symmetries of the two--dimensional plane.

As one expects, these coherent states remain localized for a short period of time (extremely short for chaotic billiards). The problem of forcing localization in the classical limit by careful choice of parameters, interaction with the environment or some other means  is a relevant problem in the broader field of quantum--classical correspondence.

\section{Acknowledgements} 
The work has been supported by HSE University Basic Research project HSE-BR-2025-6 ``Quantum devices and technologies of the new generation''.
%\bibliography{biblio}

\begin{thebibliography}{33}%
\makeatletter
\providecommand \@ifxundefined [1]{%
 \@ifx{#1\undefined}
}%
\providecommand \@ifnum [1]{%
 \ifnum #1\expandafter \@firstoftwo
 \else \expandafter \@secondoftwo
 \fi
}%
\providecommand \@ifx [1]{%
 \ifx #1\expandafter \@firstoftwo
 \else \expandafter \@secondoftwo
 \fi
}%
\providecommand \natexlab [1]{#1}%
\providecommand \enquote  [1]{``#1''}%
\providecommand \bibnamefont  [1]{#1}%
\providecommand \bibfnamefont [1]{#1}%
\providecommand \citenamefont [1]{#1}%
\providecommand \href@noop [0]{\@secondoftwo}%
\providecommand \href [0]{\begingroup \@sanitize@url \@href}%
\providecommand \@href[1]{\@@startlink{#1}\@@href}%
\providecommand \@@href[1]{\endgroup#1\@@endlink}%
\providecommand \@sanitize@url [0]{\catcode `\\12\catcode `\$12\catcode
  `\&12\catcode `\#12\catcode `\^12\catcode `\_12\catcode `\%12\relax}%
\providecommand \@@startlink[1]{}%
\providecommand \@@endlink[0]{}%
\providecommand \url  [0]{\begingroup\@sanitize@url \@url }%
\providecommand \@url [1]{\endgroup\@href {#1}{\urlprefix }}%
\providecommand \urlprefix  [0]{URL }%
\providecommand \Eprint [0]{\href }%
\providecommand \doibase [0]{https://doi.org/}%
\providecommand \selectlanguage [0]{\@gobble}%
\providecommand \bibinfo  [0]{\@secondoftwo}%
\providecommand \bibfield  [0]{\@secondoftwo}%
\providecommand \translation [1]{[#1]}%
\providecommand \BibitemOpen [0]{}%
\providecommand \bibitemStop [0]{}%
\providecommand \bibitemNoStop [0]{.\EOS\space}%
\providecommand \EOS [0]{\spacefactor3000\relax}%
\providecommand \BibitemShut  [1]{\csname bibitem#1\endcsname}%
\let\auto@bib@innerbib\@empty
%</preamble>
\bibitem [{\citenamefont {Klauder}(2001)}]{klauder_current_2001}%
  \BibitemOpen
  \bibfield  {author} {\bibinfo {author} {\bibfnamefont {J.~R.}\ \bibnamefont
  {Klauder}},\ }\href {https://doi.org/10.48550/ARXIV.QUANT-PH/0110108}
  {\bibinfo {title} {The {Current} {State} of {Coherent} {States}}} (\bibinfo
  {year} {2001})\BibitemShut {NoStop}%
\bibitem [{\citenamefont {Curado}\ \emph {et~al.}(2001)\citenamefont {Curado},
  \citenamefont {Rego-Monteiro},\ and\ \citenamefont
  {Nazareno}}]{curado_heisenberg-type_2001}%
  \BibitemOpen
  \bibfield  {author} {\bibinfo {author} {\bibfnamefont {E.~M.~F.}\
  \bibnamefont {Curado}}, \bibinfo {author} {\bibfnamefont {M.~A.}\
  \bibnamefont {Rego-Monteiro}},\ and\ \bibinfo {author} {\bibfnamefont
  {H.~N.}\ \bibnamefont {Nazareno}},\ }\bibfield  {title} {\bibinfo {title}
  {Heisenberg-type structures of one-dimensional quantum {Hamiltonians}},\
  }\href {https://doi.org/10.1103/PhysRevA.64.012105} {\bibfield  {journal}
  {\bibinfo  {journal} {Physical Review A}\ }\textbf {\bibinfo {volume} {64}},\
  \bibinfo {pages} {012105} (\bibinfo {year} {2001})}\BibitemShut {NoStop}%
\bibitem [{\citenamefont {Curado}\ \emph {et~al.}(2008)\citenamefont {Curado},
  \citenamefont {Hassouni}, \citenamefont {Rego-Monteiro},\ and\ \citenamefont
  {Rodrigues}}]{curado_generalized_2008}%
  \BibitemOpen
  \bibfield  {author} {\bibinfo {author} {\bibfnamefont {E.}~\bibnamefont
  {Curado}}, \bibinfo {author} {\bibfnamefont {Y.}~\bibnamefont {Hassouni}},
  \bibinfo {author} {\bibfnamefont {M.}~\bibnamefont {Rego-Monteiro}},\ and\
  \bibinfo {author} {\bibfnamefont {L.~M.}\ \bibnamefont {Rodrigues}},\
  }\bibfield  {title} {\bibinfo {title} {Generalized {Heisenberg} algebra and
  algebraic method: {The} example of an infinite square-well potential},\
  }\href {https://doi.org/10.1016/j.physleta.2008.01.086} {\bibfield  {journal}
  {\bibinfo  {journal} {Physics Letters A}\ }\textbf {\bibinfo {volume}
  {372}},\ \bibinfo {pages} {3350} (\bibinfo {year} {2008})}\BibitemShut
  {NoStop}%
\bibitem [{\citenamefont {Gazeau}\ and\ \citenamefont
  {Klauder}(1999)}]{gazeau_coherent_1999}%
  \BibitemOpen
  \bibfield  {author} {\bibinfo {author} {\bibfnamefont {J.~P.}\ \bibnamefont
  {Gazeau}}\ and\ \bibinfo {author} {\bibfnamefont {J.~R.}\ \bibnamefont
  {Klauder}},\ }\bibfield  {title} {\bibinfo {title} {Coherent states for
  systems with discrete and continuous spectrum},\ }\href
  {https://doi.org/10.1088/0305-4470/32/1/013} {\bibfield  {journal} {\bibinfo
  {journal} {Journal of Physics A: Mathematical and General}\ }\textbf
  {\bibinfo {volume} {32}},\ \bibinfo {pages} {123} (\bibinfo {year}
  {1999})}\BibitemShut {NoStop}%
\bibitem [{\citenamefont {Maioli}\ and\ \citenamefont
  {Curado}(2024)}]{maioli_direct_2024}%
  \BibitemOpen
  \bibfield  {author} {\bibinfo {author} {\bibfnamefont {A.~C.}\ \bibnamefont
  {Maioli}}\ and\ \bibinfo {author} {\bibfnamefont {E.~M.~F.}\ \bibnamefont
  {Curado}},\ }\href {https://doi.org/10.48550/ARXIV.2409.07385} {\bibinfo
  {title} {A direct approach to coherent states of billiards using a quantum
  algebra framework}} (\bibinfo {year} {2024})\BibitemShut {NoStop}%
\bibitem [{\citenamefont {Fox}\ and\ \citenamefont
  {Choi}(2000)}]{fox_generalized_2000}%
  \BibitemOpen
  \bibfield  {author} {\bibinfo {author} {\bibfnamefont {R.~F.}\ \bibnamefont
  {Fox}}\ and\ \bibinfo {author} {\bibfnamefont {M.~H.}\ \bibnamefont {Choi}},\
  }\bibfield  {title} {\bibinfo {title} {Generalized coherent states and
  quantum-classical correspondence},\ }\href
  {https://doi.org/10.1103/PhysRevA.61.032107} {\bibfield  {journal} {\bibinfo
  {journal} {Physical Review A}\ }\textbf {\bibinfo {volume} {61}},\ \bibinfo
  {pages} {032107} (\bibinfo {year} {2000})}\BibitemShut {NoStop}%
\bibitem [{\citenamefont {Andrews}(1998)}]{andrews_wave_1998}%
  \BibitemOpen
  \bibfield  {author} {\bibinfo {author} {\bibfnamefont {M.}~\bibnamefont
  {Andrews}},\ }\bibfield  {title} {\bibinfo {title} {Wave packets bouncing off
  walls},\ }\href {https://doi.org/10.1119/1.18854} {\bibfield  {journal}
  {\bibinfo  {journal} {American Journal of Physics}\ }\textbf {\bibinfo
  {volume} {66}},\ \bibinfo {pages} {252} (\bibinfo {year} {1998})}\BibitemShut
  {NoStop}%
\bibitem [{\citenamefont {Robinett}(2004)}]{robinett_quantum_2004}%
  \BibitemOpen
  \bibfield  {author} {\bibinfo {author} {\bibfnamefont {R.}~\bibnamefont
  {Robinett}},\ }\bibfield  {title} {\bibinfo {title} {Quantum wave packet
  revivals},\ }\href {https://doi.org/10.1016/j.physrep.2003.11.002} {\bibfield
   {journal} {\bibinfo  {journal} {Physics Reports}\ }\textbf {\bibinfo
  {volume} {392}},\ \bibinfo {pages} {1} (\bibinfo {year} {2004})}\BibitemShut
  {NoStop}%
\bibitem [{\citenamefont {Balian}\ and\ \citenamefont
  {Bloch}(1970)}]{balian_distribution_1970}%
  \BibitemOpen
  \bibfield  {author} {\bibinfo {author} {\bibfnamefont {R.}~\bibnamefont
  {Balian}}\ and\ \bibinfo {author} {\bibfnamefont {C.}~\bibnamefont {Bloch}},\
  }\bibfield  {title} {\bibinfo {title} {Distribution of eigenfrequencies for
  the wave equation in a finite domain},\ }\href
  {https://doi.org/10.1016/0003-4916(70)90497-5} {\bibfield  {journal}
  {\bibinfo  {journal} {Annals of Physics}\ }\textbf {\bibinfo {volume} {60}},\
  \bibinfo {pages} {401} (\bibinfo {year} {1970})}\BibitemShut {NoStop}%
\bibitem [{\citenamefont {Li}\ and\ \citenamefont
  {Robnik}(1995)}]{li_boundary_1995}%
  \BibitemOpen
  \bibfield  {author} {\bibinfo {author} {\bibfnamefont {B.}~\bibnamefont
  {Li}}\ and\ \bibinfo {author} {\bibfnamefont {M.}~\bibnamefont {Robnik}},\
  }\href {https://doi.org/10.48550/ARXIV.CHAO-DYN/9507002} {\bibinfo {title}
  {Boundary integral method applied in chaotic quantum billiards}} (\bibinfo
  {year} {1995})\BibitemShut {NoStop}%
\bibitem [{\citenamefont {Tiago}\ \emph {et~al.}(1997)\citenamefont {Tiago},
  \citenamefont {De~Carvalho},\ and\ \citenamefont
  {De~Aguiar}}]{tiago_boundary_1997}%
  \BibitemOpen
  \bibfield  {author} {\bibinfo {author} {\bibfnamefont {M.~L.}\ \bibnamefont
  {Tiago}}, \bibinfo {author} {\bibfnamefont {T.~O.}\ \bibnamefont
  {De~Carvalho}},\ and\ \bibinfo {author} {\bibfnamefont {M.~A.~M.}\
  \bibnamefont {De~Aguiar}},\ }\bibfield  {title} {\bibinfo {title} {Boundary
  integral method for quantum billiards in a constant magnetic field},\ }\href
  {https://doi.org/10.1103/PhysRevE.55.65} {\bibfield  {journal} {\bibinfo
  {journal} {Physical Review E}\ }\textbf {\bibinfo {volume} {55}},\ \bibinfo
  {pages} {65} (\bibinfo {year} {1997})}\BibitemShut {NoStop}%
\bibitem [{\citenamefont {Backer}\ \emph {et~al.}(2002)\citenamefont {Backer},
  \citenamefont {Furstberger}, \citenamefont {Schubert},\ and\ \citenamefont
  {Steiner}}]{backer_behaviour_2002}%
  \BibitemOpen
  \bibfield  {author} {\bibinfo {author} {\bibfnamefont {A.}~\bibnamefont
  {Backer}}, \bibinfo {author} {\bibfnamefont {S.}~\bibnamefont {Furstberger}},
  \bibinfo {author} {\bibfnamefont {R.}~\bibnamefont {Schubert}},\ and\
  \bibinfo {author} {\bibfnamefont {F.}~\bibnamefont {Steiner}},\ }\bibfield
  {title} {\bibinfo {title} {Behaviour of boundary functions for quantum
  billiards},\ }\href {https://doi.org/10.1088/0305-4470/35/48/306} {\bibfield
  {journal} {\bibinfo  {journal} {Journal of Physics A: Mathematical and
  General}\ }\textbf {\bibinfo {volume} {35}},\ \bibinfo {pages} {10293}
  (\bibinfo {year} {2002})}\BibitemShut {NoStop}%
\bibitem [{\citenamefont {Wheeler}(2000)}]{wheeler_2000}%
  \BibitemOpen
  \bibfield  {author} {\bibinfo {author} {\bibfnamefont {N.}~\bibnamefont
  {Wheeler}},\ }\href
  {https://www.reed.edu/physics/faculty/wheeler/documents/Quantum%20Mechanics/Miscellaneous%20Essays/Phase%20Space%20in%20a%20Box.pdf}
  {\bibinfo {title} {Phase space formulation of the quantum mechanical
  particle-in-a-box problem}},\ \bibinfo {howpublished} {Reed College Physics
  Department, unpublished lecture notes} (\bibinfo {year} {2000})\BibitemShut
  {NoStop}%
\bibitem [{\citenamefont {De~Vincenzo}(2008)}]{vincenzo_2008}%
  \BibitemOpen
  \bibfield  {author} {\bibinfo {author} {\bibfnamefont {S.}~\bibnamefont
  {De~Vincenzo}},\ }\bibfield  {title} {\bibinfo {title} {Impenetrable barriers
  in quantum mechanics},\ }\href@noop {} {\bibfield  {journal} {\bibinfo
  {journal} {Revista Mexicana de Fisica E}\ }\textbf {\bibinfo {volume} {54}},\
  \bibinfo {pages} {1} (\bibinfo {year} {2008})}\BibitemShut {NoStop}%
\bibitem [{\citenamefont {Al-Hashimi}\ and\ \citenamefont
  {Wiese}(2021)}]{al-hashimi_canonical_2021}%
  \BibitemOpen
  \bibfield  {author} {\bibinfo {author} {\bibfnamefont {M.~H.}\ \bibnamefont
  {Al-Hashimi}}\ and\ \bibinfo {author} {\bibfnamefont {U.-J.}\ \bibnamefont
  {Wiese}},\ }\bibfield  {title} {\bibinfo {title} {Canonical quantization on
  the half-line and in an interval based upon an alternative concept for the
  momentum in a space with boundaries},\ }\href
  {https://doi.org/10.1103/PhysRevResearch.3.033079} {\bibfield  {journal}
  {\bibinfo  {journal} {Physical Review Research}\ }\textbf {\bibinfo {volume}
  {3}},\ \bibinfo {pages} {033079} (\bibinfo {year} {2021})}\BibitemShut
  {NoStop}%
\bibitem [{\citenamefont {Seidov}(2023)}]{seidov_wigner_2023}%
  \BibitemOpen
  \bibfield  {author} {\bibinfo {author} {\bibfnamefont {S.~S.}\ \bibnamefont
  {Seidov}},\ }\bibfield  {title} {\bibinfo {title} {Wigner function dynamics
  with boundaries expressed as convolution},\ }\href
  {https://doi.org/10.1088/1751-8121/ace6e5} {\bibfield  {journal} {\bibinfo
  {journal} {Journal of Physics A: Mathematical and Theoretical}\ }\textbf
  {\bibinfo {volume} {56}},\ \bibinfo {pages} {325303} (\bibinfo {year}
  {2023})}\BibitemShut {NoStop}%
\bibitem [{\citenamefont {Seidov}\ and\ \citenamefont
  {Bezymiannykh}(2025)}]{seidov_wigner_2025}%
  \BibitemOpen
  \bibfield  {author} {\bibinfo {author} {\bibfnamefont {S.~S.}\ \bibnamefont
  {Seidov}}\ and\ \bibinfo {author} {\bibfnamefont {D.~G.}\ \bibnamefont
  {Bezymiannykh}},\ }\bibfield  {title} {\bibinfo {title} {Wigner current in
  multidimensional quantum billiards},\ }\href
  {https://doi.org/10.1088/1751-8121/ad9dc4} {\bibfield  {journal} {\bibinfo
  {journal} {Journal of Physics A: Mathematical and Theoretical}\ }\textbf
  {\bibinfo {volume} {58}},\ \bibinfo {pages} {025301} (\bibinfo {year}
  {2025})}\BibitemShut {NoStop}%
\bibitem [{\citenamefont {Burkov}\ and\ \citenamefont
  {Seidov}(2025)}]{burkov_features_2025}%
  \BibitemOpen
  \bibfield  {author} {\bibinfo {author} {\bibfnamefont {I.~D.}\ \bibnamefont
  {Burkov}}\ and\ \bibinfo {author} {\bibfnamefont {S.~S.}\ \bibnamefont
  {Seidov}},\ }\bibfield  {title} {\bibinfo {title} {Features of {Trajectories}
  {Formed} by the {Wigner} {Current}},\ }\href
  {https://doi.org/10.1134/S1062873825713212} {\bibfield  {journal} {\bibinfo
  {journal} {Bulletin of the Russian Academy of Sciences: Physics}\ }\textbf
  {\bibinfo {volume} {89}},\ \bibinfo {pages} {2088} (\bibinfo {year}
  {2025})}\BibitemShut {NoStop}%
\bibitem [{\citenamefont {Lange}(2012)}]{lange_potential_2012}%
  \BibitemOpen
  \bibfield  {author} {\bibinfo {author} {\bibfnamefont {R.-J.}\ \bibnamefont
  {Lange}},\ }\bibfield  {title} {\bibinfo {title} {Potential theory, path
  integrals and the {Laplacian} of the indicator},\ }\href
  {https://doi.org/10.1007/JHEP11(2012)032} {\bibfield  {journal} {\bibinfo
  {journal} {Journal of High Energy Physics}\ }\textbf {\bibinfo {volume}
  {2012}},\ \bibinfo {pages} {32} (\bibinfo {year} {2012})}\BibitemShut
  {NoStop}%
\bibitem [{\citenamefont {Lange}(2015)}]{lange_distribution_2015}%
  \BibitemOpen
  \bibfield  {author} {\bibinfo {author} {\bibfnamefont {R.-J.}\ \bibnamefont
  {Lange}},\ }\bibfield  {title} {\bibinfo {title} {Distribution theory for
  {Schrödinger}’s integral equation},\ }\href
  {https://doi.org/10.1063/1.4936302} {\bibfield  {journal} {\bibinfo
  {journal} {Journal of Mathematical Physics}\ }\textbf {\bibinfo {volume}
  {56}},\ \bibinfo {pages} {122105} (\bibinfo {year} {2015})}\BibitemShut
  {NoStop}%
\bibitem [{\citenamefont {Dias}\ and\ \citenamefont
  {Prata}(2002)}]{dias_wigner_2002}%
  \BibitemOpen
  \bibfield  {author} {\bibinfo {author} {\bibfnamefont {N.~C.}\ \bibnamefont
  {Dias}}\ and\ \bibinfo {author} {\bibfnamefont {J.~N.}\ \bibnamefont
  {Prata}},\ }\bibfield  {title} {\bibinfo {title} {Wigner functions with
  boundaries},\ }\href {https://doi.org/10.1063/1.1504885} {\bibfield
  {journal} {\bibinfo  {journal} {Journal of Mathematical Physics}\ }\textbf
  {\bibinfo {volume} {43}},\ \bibinfo {pages} {4602} (\bibinfo {year}
  {2002})}\BibitemShut {NoStop}%
\bibitem [{\citenamefont {Dias}\ and\ \citenamefont
  {Prata}(2021)}]{dias_boundaries_2021}%
  \BibitemOpen
  \bibfield  {author} {\bibinfo {author} {\bibfnamefont {N.~C.}\ \bibnamefont
  {Dias}}\ and\ \bibinfo {author} {\bibfnamefont {J.~N.}\ \bibnamefont
  {Prata}},\ }\bibfield  {title} {\bibinfo {title} {Boundaries and profiles in
  the {Wigner} formalism},\ }\href {https://doi.org/10.1007/s10825-021-01803-7}
  {\bibfield  {journal} {\bibinfo  {journal} {Journal of Computational
  Electronics}\ }\textbf {\bibinfo {volume} {20}},\ \bibinfo {pages} {2020}
  (\bibinfo {year} {2021})}\BibitemShut {NoStop}%
\bibitem [{\citenamefont {Maioli}\ and\ \citenamefont
  {Schmidt}(2018)}]{maioli_exact_2018}%
  \BibitemOpen
  \bibfield  {author} {\bibinfo {author} {\bibfnamefont {A.~C.}\ \bibnamefont
  {Maioli}}\ and\ \bibinfo {author} {\bibfnamefont {A.~G.~M.}\ \bibnamefont
  {Schmidt}},\ }\bibfield  {title} {\bibinfo {title} {Exact solution to
  {Lippmann}-{Schwinger} equation for a circular billiard},\ }\href
  {https://doi.org/10.1063/1.5056259} {\bibfield  {journal} {\bibinfo
  {journal} {Journal of Mathematical Physics}\ }\textbf {\bibinfo {volume}
  {59}},\ \bibinfo {pages} {122102} (\bibinfo {year} {2018})}\BibitemShut
  {NoStop}%
\bibitem [{\citenamefont {Gori}\ \emph {et~al.}(2001)\citenamefont {Gori},
  \citenamefont {Ambrosini}, \citenamefont {Borghi}, \citenamefont {Mussi},\
  and\ \citenamefont {Santarsiero}}]{gori_propagator_2001}%
  \BibitemOpen
  \bibfield  {author} {\bibinfo {author} {\bibfnamefont {F.}~\bibnamefont
  {Gori}}, \bibinfo {author} {\bibfnamefont {D.}~\bibnamefont {Ambrosini}},
  \bibinfo {author} {\bibfnamefont {R.}~\bibnamefont {Borghi}}, \bibinfo
  {author} {\bibfnamefont {V.}~\bibnamefont {Mussi}},\ and\ \bibinfo {author}
  {\bibfnamefont {M.}~\bibnamefont {Santarsiero}},\ }\bibfield  {title}
  {\bibinfo {title} {The propagator for a particle in a well},\ }\href
  {https://doi.org/10.1088/0143-0807/22/1/306} {\bibfield  {journal} {\bibinfo
  {journal} {European Journal of Physics}\ }\textbf {\bibinfo {volume} {22}},\
  \bibinfo {pages} {53} (\bibinfo {year} {2001})}\BibitemShut {NoStop}%
\bibitem [{\citenamefont {Doncheski}\ \emph {et~al.}(2003)\citenamefont
  {Doncheski}, \citenamefont {Heppelmann}, \citenamefont {Robinett},\ and\
  \citenamefont {Tussey}}]{doncheski_wave_2003}%
  \BibitemOpen
  \bibfield  {author} {\bibinfo {author} {\bibfnamefont {M.~A.}\ \bibnamefont
  {Doncheski}}, \bibinfo {author} {\bibfnamefont {S.}~\bibnamefont
  {Heppelmann}}, \bibinfo {author} {\bibfnamefont {R.~W.}\ \bibnamefont
  {Robinett}},\ and\ \bibinfo {author} {\bibfnamefont {D.~C.}\ \bibnamefont
  {Tussey}},\ }\bibfield  {title} {\bibinfo {title} {Wave packet construction
  in two-dimensional quantum billiards: {Blueprints} for the square,
  equilateral triangle, and circular cases},\ }\href
  {https://doi.org/10.1119/1.1538574} {\bibfield  {journal} {\bibinfo
  {journal} {American Journal of Physics}\ }\textbf {\bibinfo {volume} {71}},\
  \bibinfo {pages} {541} (\bibinfo {year} {2003})}\BibitemShut {NoStop}%
\bibitem [{\citenamefont {Dubrovin}(1981)}]{dubrovin_theta_1981}%
  \BibitemOpen
  \bibfield  {author} {\bibinfo {author} {\bibfnamefont {B.~A.}\ \bibnamefont
  {Dubrovin}},\ }\bibfield  {title} {\bibinfo {title} {Theta functions and
  non-linear equations},\ }\href
  {https://doi.org/10.1070/RM1981v036n02ABEH002596} {\bibfield  {journal}
  {\bibinfo  {journal} {Russian Mathematical Surveys}\ }\textbf {\bibinfo
  {volume} {36}},\ \bibinfo {pages} {11} (\bibinfo {year} {1981})}\BibitemShut
  {NoStop}%
\bibitem [{\citenamefont {Deconinck}\ \emph {et~al.}(2002)\citenamefont
  {Deconinck}, \citenamefont {Heil}, \citenamefont {Bobenko}, \citenamefont
  {Hoeij},\ and\ \citenamefont {Schmies}}]{deconinck_computing_2002}%
  \BibitemOpen
  \bibfield  {author} {\bibinfo {author} {\bibfnamefont {B.}~\bibnamefont
  {Deconinck}}, \bibinfo {author} {\bibfnamefont {M.}~\bibnamefont {Heil}},
  \bibinfo {author} {\bibfnamefont {A.}~\bibnamefont {Bobenko}}, \bibinfo
  {author} {\bibfnamefont {M.~v.}\ \bibnamefont {Hoeij}},\ and\ \bibinfo
  {author} {\bibfnamefont {M.}~\bibnamefont {Schmies}},\ }\href
  {https://doi.org/10.48550/arXiv.nlin/0206009} {\bibinfo {title} {Computing
  {Riemann} {Theta} {Functions}}} (\bibinfo {year} {2002}),\ \bibinfo {note}
  {arXiv:nlin/0206009}\BibitemShut {NoStop}%
\bibitem [{\citenamefont
  {Wheeler}(1997{\natexlab{a}})}]{wheeler1997appliedtheta}%
  \BibitemOpen
  \bibfield  {author} {\bibinfo {author} {\bibfnamefont {N.}~\bibnamefont
  {Wheeler}},\ }\href
  {https://www.reed.edu/physics/faculty/wheeler/documents/Quantum%20Mechanics/Miscellaneous%20Essays/Applied%20Theta%20Workshop.pdf}
  {\bibinfo {title} {Applied theta functions of one or several variables}}
  (\bibinfo {year} {1997}{\natexlab{a}}),\ \bibinfo {note} {reed College
  Physics Department; prepared for the Fall Semester Series of faculty seminars
  organized by the Reed College Mathematics Department}\BibitemShut {NoStop}%
\bibitem [{\citenamefont
  {Wheeler}(1997{\natexlab{b}})}]{wheeler1997particlebox}%
  \BibitemOpen
  \bibfield  {author} {\bibinfo {author} {\bibfnamefont {N.}~\bibnamefont
  {Wheeler}},\ }\href
  {https://www.reed.edu/physics/faculty/wheeler/documents/Quantum%20Mechanics/Miscellaneous%20Essays/2D%20Box%20Problems.pdf}
  {\bibinfo {title} {2-dimensional ``particle-in-a-box'' problems in quantum
  mechanics}} (\bibinfo {year} {1997}{\natexlab{b}}),\ \bibinfo {note} {reed
  College Physics Department}\BibitemShut {NoStop}%
\bibitem [{\citenamefont {Turner}(1984)}]{turner_quantum_1984}%
  \BibitemOpen
  \bibfield  {author} {\bibinfo {author} {\bibfnamefont {J.~W.}\ \bibnamefont
  {Turner}},\ }\bibfield  {title} {\bibinfo {title} {On the quantum particle in
  a polyhedral box},\ }\href {https://doi.org/10.1088/0305-4470/17/14/022}
  {\bibfield  {journal} {\bibinfo  {journal} {Journal of Physics A:
  Mathematical and General}\ }\textbf {\bibinfo {volume} {17}},\ \bibinfo
  {pages} {2791} (\bibinfo {year} {1984})}\BibitemShut {NoStop}%
\bibitem [{\citenamefont {Terras}\ and\ \citenamefont
  {Swanson}(1980)}]{terras_image_1980}%
  \BibitemOpen
  \bibfield  {author} {\bibinfo {author} {\bibfnamefont {R.}~\bibnamefont
  {Terras}}\ and\ \bibinfo {author} {\bibfnamefont {R.}~\bibnamefont
  {Swanson}},\ }\bibfield  {title} {\bibinfo {title} {Image methods for
  constructing {Green}’s functions and eigenfunctions for domains with plane
  boundaries},\ }\href {https://doi.org/10.1063/1.524723} {\bibfield  {journal}
  {\bibinfo  {journal} {Journal of Mathematical Physics}\ }\textbf {\bibinfo
  {volume} {21}},\ \bibinfo {pages} {2140} (\bibinfo {year}
  {1980})}\BibitemShut {NoStop}%
\bibitem [{\citenamefont {Li}\ and\ \citenamefont
  {Blinder}(1987)}]{li_particle_1987}%
  \BibitemOpen
  \bibfield  {author} {\bibinfo {author} {\bibfnamefont {W.-K.}\ \bibnamefont
  {Li}}\ and\ \bibinfo {author} {\bibfnamefont {S.~M.}\ \bibnamefont
  {Blinder}},\ }\bibfield  {title} {\bibinfo {title} {Particle in an
  equilateral triangle: {Exact} solution of a nonseparable problem},\ }\href
  {https://doi.org/10.1021/ed064p130} {\bibfield  {journal} {\bibinfo
  {journal} {Journal of Chemical Education}\ }\textbf {\bibinfo {volume}
  {64}},\ \bibinfo {pages} {130} (\bibinfo {year} {1987})}\BibitemShut
  {NoStop}%
\bibitem [{\citenamefont {Erman}\ \emph {et~al.}(2020)\citenamefont {Erman},
  \citenamefont {Gadella},\ and\ \citenamefont
  {Uncu}}]{erman_propagators_2020}%
  \BibitemOpen
  \bibfield  {author} {\bibinfo {author} {\bibfnamefont {F.}~\bibnamefont
  {Erman}}, \bibinfo {author} {\bibfnamefont {M.}~\bibnamefont {Gadella}},\
  and\ \bibinfo {author} {\bibfnamefont {H.}~\bibnamefont {Uncu}},\ }\bibfield
  {title} {\bibinfo {title} {The {Propagators} for $\delta$ and $\delta'$
  {Potentials} {With} {Time}-{Dependent} {Strengths}},\ }\href
  {https://doi.org/10.3389/fphy.2020.00065} {\bibfield  {journal} {\bibinfo
  {journal} {Frontiers in Physics}\ }\textbf {\bibinfo {volume} {8}},\ \bibinfo
  {pages} {65} (\bibinfo {year} {2020})}\BibitemShut {NoStop}%
\end{thebibliography}

\appendix
\section{Connection to $\delta'_{\partial \Omega}(r)$--potential}
The time--dependent Schroedinger equation, corresponding to the Hamiltonian (\ref{eq:H_delta}), is
\begin{equation}
i \dot \psi(r, t) = -\frac{1}{2 m} \psi''(r, t) + \frac{1}{2 m}\delta'_{\partial \Omega}(r) \psi(r).
\end{equation}
In spirit of \cite{erman_propagators_2020}, where the one--dimensional delta and delta--prime potentials were considered, we perform the Laplace transform with respect to time and the Fourier transform with respect to spatial coordinate. This gives the transformed equation
\begin{equation}
\hat{\tilde{\psi}}(p, s) = \frac{1}{i s - p^2/(2 m)} \left[i \hat\psi(p, 0) + \frac{1}{2 m}\hat\delta'_{\partial B}(p) *_p \hat{\tilde{\psi}}(p, s) \right].
\end{equation}
The first term in the right hand side is the transform of the free particle wave function:
\begin{equation}
\begin{aligned}
&\mathcal{L}^{-1}_{s \rightarrow t} \mathcal{F}^{-1}_{p \rightarrow r} \left\{\frac{i \hat\psi(p, 0)}{i s - p^2/(2 m)}\right\} = \mathcal{G}_0(r, t) *_r \psi(r, 0)\\
&\frac{i}{i s - p^2/(2 m)} = \hat{\tilde{\mathcal{G}}}_0(p, s) .
\end{aligned}
\end{equation}
This brings the equation to form
\begin{equation}\label{eq:psi_ps}
\hat{\tilde{\psi}}(p, s) = \hat{\tilde{\psi}}_0(p, s) - \frac{i}{2m} \hat{\tilde{\mathcal{G}}}_0(p, s) \left[ \hat\delta'_{\partial \Omega}(p) *_p \hat{\tilde{\psi}}(p, s) \right].
\end{equation}
Now by transforming back to spatial coordinate via the inverse Fourier transform we find
\begin{equation}
\tilde \psi(r, s) = \tilde \psi_0(r, s) - \frac{i}{2m} \tilde{\mathcal{G}}_0(r, s) *_r \left[\delta'_{\partial \Omega} \tilde \psi(r, s) \right].
\end{equation}
The convolution in the second term can be brought into the form of a line integral over the boundary using (\ref{eq:delta_s}) as follows:
\begin{equation}
\begin{aligned}
\tilde{\mathcal{G}}_0(r, s) *_r \left[\delta'_{\partial \Omega} \tilde \psi(r, s) \right] &= \int \mathcal{G}_0(r - r', s) \delta'_{\partial \Omega} (r') \tilde \psi(r', s) d \lambda_{r'} =\\ 
&= \oint\limits_{\partial \Omega} \partial_{n(r')} \left[\tilde{\mathcal{G}}_0(r - r', s) \tilde \psi(r', s) \right] d \lambda_{r'} =\\
&= \oint\limits_{\partial \Omega} \left[\partial_{n(r')} \tilde{\mathcal{G}}_0(r - r', s) \tilde \psi(r', s) +  \tilde{\mathcal{G}}_0(r - r', s) \partial_{n(r')} \tilde \psi(r', s) \right] d \lambda_{r'}.
\end{aligned}
\end{equation}
The Dirichlet boundary conditions require $\tilde \psi(r \in \partial \Omega, s) = 0$, so only the second term in the integral is left. This finally leads to equation
\begin{equation}\label{eq:psi_rs}
\tilde \psi(r, s) = \tilde \psi_0(r, s) - \frac{i}{2 m} \oint\limits_{\partial \Omega} \tilde{\mathcal{G}_0}(r - r', s) \partial_{n(r')} \tilde \psi(r', s) d \lambda_{r'}.
\end{equation}
After performing the inverse Laplace transform with respect to the variable $s$ the Balian--Bloch equation (\ref{eq:BB}) is obtained.

\end{document}